# FEASIBILITY OF DIAGNOSTICS UNDULATOR STUDIES AT ASTA*

A.H. Lumpkin[#] and M. Wendt, Fermi National Accelerator Laboratory, Batavia, IL 60510 USA
J.M. Byrd, Lawrence Berkeley National Laboratory, Berkeley, CA 94720 USA

*Abstract*

The Advanced Superconducting Test Accelerator (ASTA) facility is currently under construction at Fermilab. With a 1-ms macropulse composed of up to 3000 micropulses and with beam energies projected from 45 to 800 MeV, the need for non-intercepting diagnostics for beam size, position, energy, and bunch length is clear. In addition to the rf BPMs, optical synchrotron radiation (OSR), and optical diffraction radiation (ODR) techniques already planned, we propose the use of undulator radiation from a dedicated device for diagnostics. with a nominal period of 4-5 cm, a tunable field parameter $K$, and a length of several meters. The feasibility of extending such techniques in the visible regime at a beam energy of 125 MeV into the UV and VUV regimes with beam energies of 250 and 500 MeV will be presented.

## INTRODUCTION

One of the challenges of the present-day and proposed superconducting linear accelerators with concomitant high-power beams is the non-intercepting diagnostics of the beam size, energy, bunch length, and phase. The acquisition of a comprehensive set of electron beam properties from a linear accelerator based on undulator radiation emitted from a 5-m long device was demonstrated over two decades ago on a visible wavelength free-electron laser (FEL) [1] driven by a pulse train of 100-μs extent. The high-power electron beams for the Advanced Superconducting Test Accelerator (ASTA) facility involve up to 3000 micropulses with up to 3.2 nC per micropulse in a 1-ms macropulse [2]. With beam energies projected from 45 to 800 MeV the need for non-intercepting diagnostics is clear. Besides the rf BPMs, optical synchrotron radiation (OSR), and optical diffraction radiation (ODR) techniques already considered, we propose the use of the properties of undulator radiation (UR) from a dedicated device for diagnostics with a nominal period of 4-5 cm, a tunable field parameter $K$, and a length of several meters. We propose time resolving the e-beam properties within the macropulse by viewing the UR with standard electronic-shuttered CCDs or gated ICCD's (size and position) and a synchroscan streak camera coupled to an optical spectrometer (energy, bunch length, and phase).

Initial tests could begin at a beam energy of 100-125 MeV with UR in the visible regime and could be extended into the UV and VUV regimes with beam energies of 250 and 500 MeV.

## FACILITY ASPECTS

The ASTA linac with photocathode (PC) rf gun, two booster L-band SCRF accelerators (CC1 and CC2), and beamline is schematically shown in Fig. 1. The L-band accelerating sections will provide 40- to 50-MeV beams before the chicane, and an additional acceleration capability up to a total of 800 MeV will eventually be installed in the form of three cryomodules after the chicane with eight 9-cell cavities with highest possible average gradient (up to ~30 MV/m). The phase of the CC2 section can be adjusted to energy chirp the beam entering the chicane to vary bunch-length compression. Maximizing the far infrared (FIR) coherent transition radiation (CTR) in a detector after the chicane will be used as the signature of generating the shortest bunch lengths. Micropulse charges of 20 to 3200 pC will be used typically as indicated in Table 1. The nominal micropulse format is 3 MHz for 1 ms. This aspect is unique for test facilities in the USA and highly relevant to the next generation of FELs. The macropulse repetition rate will be 5 Hz.

Signal strengths should allow tracking of a subset of the micropulses with standard imaging. The UR pulse length will be measured with the Hamamatsu UV-visible C5680 streak camera, and this length should be correlated with the electron beam bunch length subject to some narrowing when there is SASE gain. The synchroscan streak camera will also allow tracking of the relative phase within the macropulse of sets of micropulses to about 200 fs. At this time we anticipate one would provide optical transport for the signals to the high-energy -end laser lab where a small diagnostics suite of CCD camera, ICCD camera, streak camera, and optical spectrometer would be available on an optics table for characterizing the UR properties and then the deduced

Table 1: Summary of Electron Beam Properties at ASTA

| Parameter | Units | Values |
| --- | --- | --- |
| Bunch charge | pC | 20-3200 |
| Emittance, norm | mm mrad | 1-3 |
| Bunch length, rms | ps | 3-1 |
| Micropulse Number | | 1-3000 |

___________________
[#] lumpkin@fnal.gov
*Work supported under Contract No. DE-AC02-07CH11359 with the United States Department of Energy.

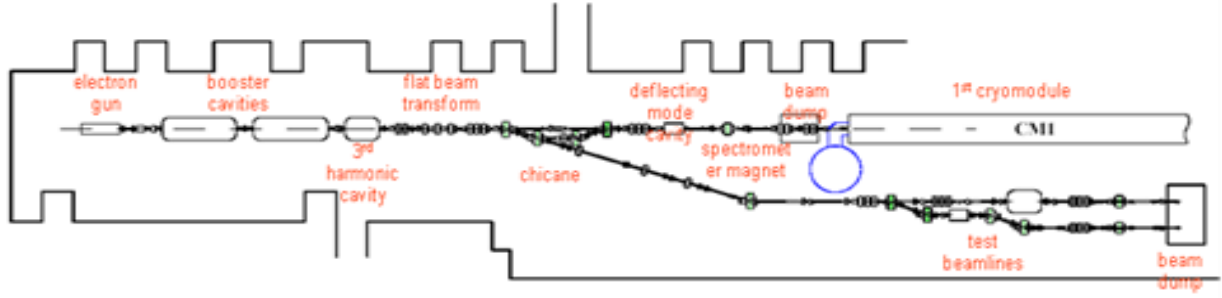

Figure 1: Schematic of the injector for the ASTA facility showing PC rf gun, booster accelerators, and beamlines. (courtesy of M. Church).

electron beam properties. Initial detection could be in the tunnel with local camera stations.

## CONCEPTUAL ASPECTS of UR

The propagation of the electron beam through the alternating static magnetic fields of an undulator results in the generation of photons. This is initiated through the spontaneous emission radiation (SER) process, but under resonance conditions a favorable instability evolves as the electron beam co-propagates with the photon fields and the electron beam is microbunched at the resonant wavelength leading to a self-amplified spontaneous emission (SASE) free-electron laser (FEL). For a planar undulator, the radiation generation process on axis is governed by the resonance condition:

$$\lambda = \lambda_u (1 + K^2/2)/2n\gamma^2, \qquad \text{Eq. 1}$$

where $\lambda$ is the UR wavelength, $\lambda_u$ is the undulator period, $K$ is the undulator field strength parameter, n is the harmonic number, and $\gamma$ is the relativistic Lorentz factor [3].

We will use the UR properties for deducing time–resolved electron beam properties. The base image size and position will be seen in the UR spot size and transverse position (although the size will have depth-of-focus issues to address). A telescope with limited depth of focus may be used to emphasize a shorter z length within the undulator's 4.5 m length. The central wavelength of the emission is directly correlated with electron-beam energy, and the bandwidth from the UR will be about 1/nN, where N is the number of periods, or 1% for 100 periods in the fundamental.

## PROPOSED UR STUDIES

It is proposed to start the UR studies with the commissioning of the first cryomodule at a beam energy of 125 MeV. As shown in Table 2 with a 5.0-cm period undulator and the gap adjusted for $K$=1.2, the fundamental UR would be at 680 nm with a third harmonic at 226 nm. One could track the beam at higher energies by reducing the gap and increasing the $K$ value so that the resonance condition remains in the visible–UV regime to simplify the early tests. One can also consider 150-, 200-, and 250-MeV cases as given in the table. The Phase numbers 1-4 are basically related to the number of installed cryomodules with up to 250 MeV beam accelerating capability per cryomodule.

Table 2: Summary of possible SER/UR wavelengths generated with a 5.0-cm period undulator at ASTA starting at 125-MeV e-beam energy. The last case is for a SC undulator with shorter period built in England for the ILC R&D program [4].

| Phase No. | Beam Energy (MeV) | Und. Fundamental (nm) | Period (cm), $K$ | Undulator Radiation Harmonics (nm) 3, 5 |
|---|---|---|---|---|
| 1 | 125 | 680 | 5.0,1.2 | 226 |
| 1 | 150 | 472 | 5.0,1.2 | 157 |
|   | 200 | 265 | 5.0,1.2 | 88 |
| 1 | 250 | 170 | 5.0,1.2 | 57 |
| 1 | 250 | 262 | 5.0,1.8 | 87 |
| 2 | 500 | 42 | 5.0,1.2 | 14, 8.3 |
| 3 | 800 | 13.4 | 5.0,0.9 | 4.5 |
| 4 | 900 | 3 | 1.1,0.9 | -- |

At 150 MeV the beam will be shared with the injection task for the Integrable Optics Test Accelerator (IOTA) tests in a parasitic mode. The configuration is shown in Fig. 2 (top view) where the diagnostics undulator is in the straight-ahead AARD area, and IOTA is reached through the transport line to the right. At these wavelengths and for the expected beam quality the gain length should be sub-m so sufficient SASE-induced microbunching should also occur in the 4.5-m length that could be accessed through coherent optical transition radiation (COTR) techniques [5,6]. In the past these have allowed evaluation and adjusting of the critical electron-beam-photon-beam overlap.

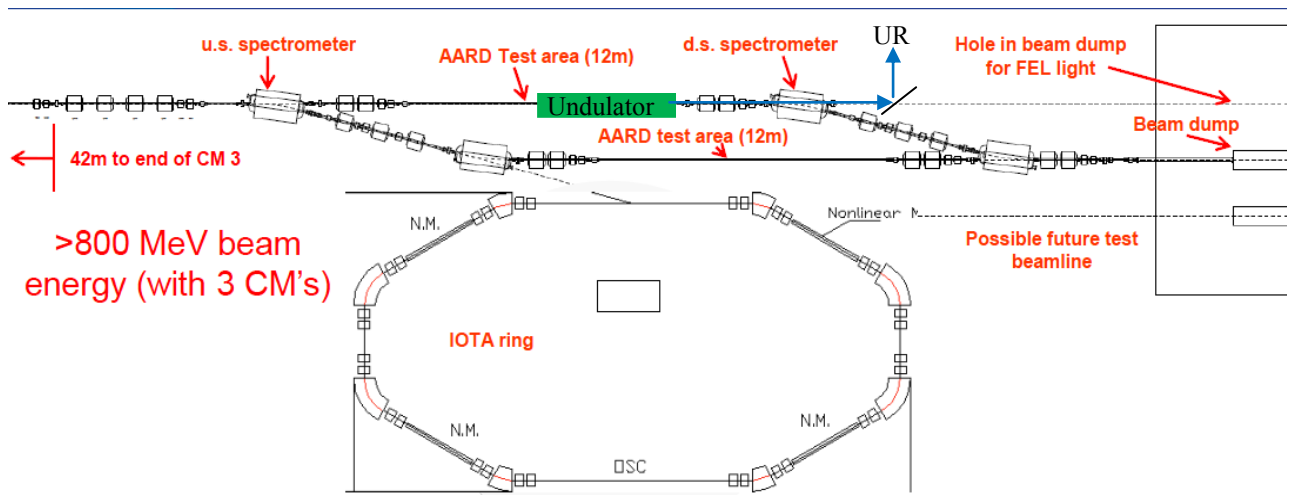

Figure 2: Schematic of the diagnostics undulator located in one high energy test area at ASTA. The output UR is accessed with an in-vacuum mirror/transport after the next dipole. The location of the IOTA ring is also indicated. (courtesy of M. Church, revised).

## VUV-XUV FEL OSCILLATOR

Following single-pass tests of beam through the undulator, the 3-MHz micropulse repetition rate for 1 ms could enable FEL oscillator tests over a broad range of wavelengths in the vacuum ultraviolet (VUV) and for the first time to the extreme ultraviolet (XUV). Rough estimates using on-axis cavity mirrors of $MgF_2$ coated Al for wavelengths from 180 to 120 nm show promise. In addition, preliminary calculations by M. Reinsch (LBNL) using both a 1-D oscillator code and the GINGER 3-D simulation code for cases at ~ 13.4 nm show feasiblity [7] when invoking the new generation of multilayer metal mirrors with 68% reflectance at 90 degree incidence to the surface [8]. The FEL output results calculated by GINGER for 800 A peak current, 4.5-m undulator length, and the base e-beam parameters in Table 1 are shown in Fig. 3. Saturation is reached after 300 passes with potentially another 900 μsec at that level subject to heat loads. Other optical resonator configurations and mirror cooling may be needed to address the heat loading and consequent thermal distortions of the mirror surfaces.

## PRACTICAL CONSIDERATIONS

There is an option for the U5.0 device [9] that is being retired from the ALS storage ring in January 2013 to be transferred to FNAL. There are some practical considerations, however, mostly driven by the labeled weight of 47,000 lbs which is close to the ASTA facility crane's 25-ton capacity. The undulator has a 5.0-cm period, 4.5-m magnetic structure length, a maximum field strength of 0.87 T with 1.4-cm gap, and a remotely controlled adjustable gap from 1.4 to 21.4 cm leading to a range of K values. The strong back and the specification to have the device's gap constant to about 58 μm over the 4.5-m length leads to the very substantial support structure. Its total height is 2.4 m as compared to the

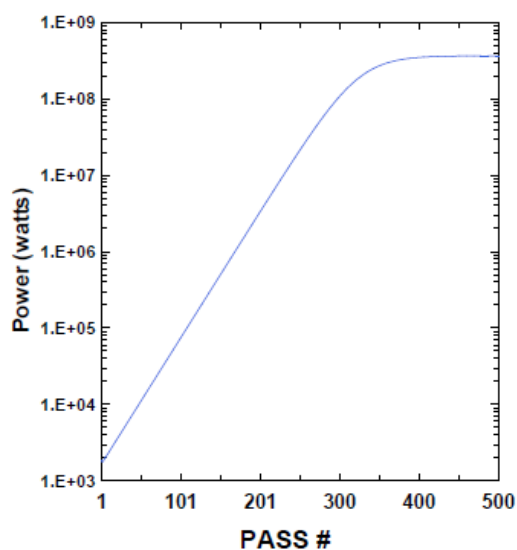

Figure 3: Initial GINGER simulations for output power saturation at the fundamental 13.4-nm wavelength for a concentric cavity with multilayer mirrors and peak current of 800 A at 800 MeV. The output power growth with pass number is shown. (courtesy of M. Reinsch [7]).

ASTA tunnel height of 3.2 m (10.5 ft). The vacuum chamber is 5.1 m flange to flange and has an antechamber. As staged at ALS, the beamline centerline is at ~55 inches, higher than the ASTA beamline center of 48 inches. There appears in reference [9] to be a baseplate under the frame of about 15 in. height that might be shortened. These should be evaluated for modification, or we have to shift the beam trajectory upward with a dogleg in this area. A photograph of the device in situ is shown in Fig. 4. Initial device characterizations were previously reported [10].

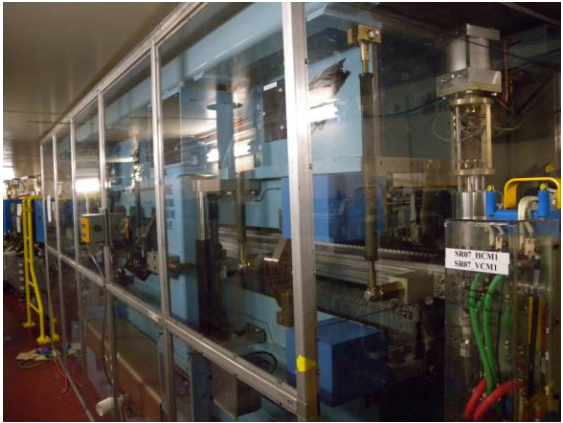

Figure 4: Photograph of the ALS U5.0 Undulator in the storage ring in April 2012. (photo by M. Wendt).

## SUMMARY

In summary, we have described a possible application at ASTA of a 5-cm period undulator with 4.5-m length for non-intercepting e-beam diagnostics during beam commissioning of the first cryomodule and subsequent installations of the other cryomodules. In addition, the undulator could be the basis of unique tests of VUV-XUV FEL oscillator configurations. A more comprehensive study of this latter topic is underway.

## ACKNOWLEDGMENTS

The FNAL authors acknowledge discussions with M. Church, V. Shiltsev, S. Nagaitsev, and S. Henderson of FNAL on the ASTA facility and AARD program and all authors acknowledge M. Reinsch of LBNL for providing the initial FEL oscillator simulations.